# An Analytical Model of Sorption-Induced Static Mode Nanomechanical Sensing for Multi-Component Analytes

Kosuke Minami*[,†,‡] and Genki Yoshikawa[†,§]

[†] Research Center for Macromolecules and Biomaterials, National Institute for Materials Science (NIMS), 1-1 Namiki, Tsukuba, Ibaraki 305-0044 Japan

[‡] International Center for Young Scientists (ICYS), National Institute for Materials Science (NIMS), 1-1 Namiki, Tsukuba, Ibaraki 305-0044 Japan

[§] Materials Science and Engineering, Graduate School of Pure and Applied Science, University of Tsukuba, 1-1-1 Tennodai, Tsukuba, Ibaraki 305-8571, Japan

* Correspondence and requests for materials should be addressed to K.M. (MINAMI.Kosuke@nims.go.jp)

**ABSTRACT:** Nanomechanical sensors and their arrays have attracted significant attention for detecting, distinguishing, and identifying target analytes, especially complex mixtures of odors. In the static mode operation, sensing signals are obtained by a concentration-dependent sorption-induced mechanical strain/stress. Understanding of the dynamic responses is crucial for developing practical artificial olfaction; however, the analytical formulations are still limited to single-component analytes. Here, we derive an analytical model of viscoelastic material-based static mode nanomechanical sensing for multi-component analytes by extending the theoretical model via solving differential equations. The present model can reduce the dynamic responses to the multi-component target analytes observed in the experimental signal responses. Moreover, the use of optimized fitting parameters extracted from pure vapors with viscoelastic parameters allows us to predict the concentration of each analyte in the multi-component system.

Detecting odors composed of complex mixtures of gaseous molecules is a fundamental requirement for a wide range of applications in electronic noses—models of the nose using an array of chemical sensors. Since its proposal by Persaud and Dodd,[1] electronic noses and related technologies have been extensively studied.[2-8] Understanding the dynamic responses to gases is critically important not only to analyze target analytes in the practical samples[9,10] but also to extract effective features for subsequent multivariate analyses, including machine learning.[11-14] However, analytical investigation of the dynamic responses of chemical sensors is still limited.

Among the various sensors reported, nanomechanical sensors have attracted much attention, partly because almost all materials can be used as a receptor layer.[5,15,16] Thus, an array of nanomechanical sensors can be potentially used as a sensing unit for the detection of the complex mixtures of odors in various fields,[5] including food,[11,13,17] agriculture,[9,10,18,19] environment,[20-22] and medical and healthcare fields.[23-29] For the so-called static mode operation of nanomechanical sensors,[5,30] the sensing signals are obtained via deformation induced by the sorption of target molecules in a receptor layer (Figure 1a,b). Such sorption behavior in static mode nanomechanical sensing is theoretically investigated for elastic and viscoelastic coatings.[31-34] However, those theoretical formulations are derived from the models of single analyte absorption and are still unable to be applied to multi-component systems such as complex mixtures of odors.

In this study, we extend the theoretical studies based on single component systems[32,33] to multi-component systems and propose a general expression that includes sorption kinetics along with viscoelastic stress relaxation of receptor materials for nanomechanical sensing in static mode operation. We formulated a new model by driving an analytical solution of overall transient responses to the multi-component analytes. This analytical model agrees well with sensor responses to the vapors of binary mixtures experimentally measured using one of the nanomechanical sensors in static mode operation—Membrane-type Surface stress Sensor (MSS)[35,36]—coated with viscoelastic receptor materials (Figure 1b). The present model can be utilized not only for reproducing entire dynamic responses of nanomechanical sensors but also for analyzing multi-component vapors.

## EXPERIMENTAL SECTION

**Materials.** Polycaprolactone (PCL) and poly(vinylidene fluoride) (PVF) were purchased from Sigma-Aldrich Inc. *N,N*-Dimethylformamide (DMF) was purchased from Sigma-Aldrich Inc. and used for preparation of receptor layers. *n*-Hexane, *n*-nonane, *n*-dodecane, methanol, ethanol, and 2-propanol were purchased Sigma-Aldrich Inc., Tokyo Chemical Industry Co., Inc., or Nacalai Tesque Co., Ltd.

**Sensor Preparation.** The construction of the MSS chips and its working principle have been previously reported (Figure 1a,b).[35,36] Briefly, MSS consists of a silicon membrane suspended by four sensing beams, in which piezoresistors are embedded. The membrane is coated with a receptor material, which generates surface stress caused by the sorption-induced expansion. PCL and PVF were deposited directly onto the MSS membrane by inkjet spotting using an inkjet spotter (LaboJet-500SP, MICROJET Co. Ltd.) equipped with a nozzle (IJHBS-300, MICROJET Co. Ltd.). Each polymer was dissolved in DMF at a concentration of 1 mg/mL, and the resulting solutions were deposited onto each surface of the MSS membrane. The inkjet conditions are used according to the previous study.[28]

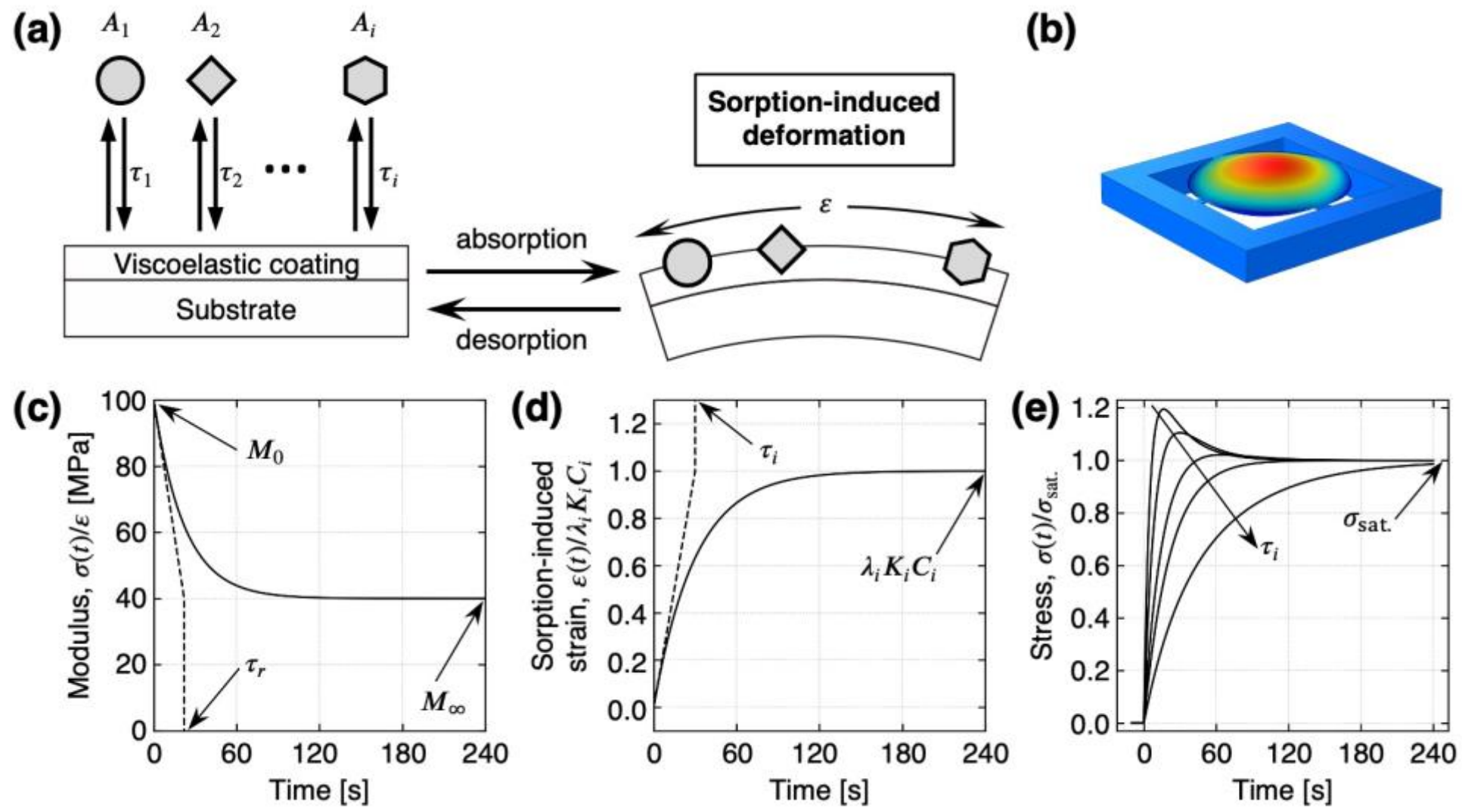

**Figure 1.** Sorption-induced model for nanomechanical sensors. **a)** Working principle of nanomechanical sensors in static mode operation. **b)** Typical geometries of one of the nanomechanical sensors, MSS. Color gradient represents the displacement in *z*-direction (i.e., perpendicular to the cantilever surface or membrane surface) simulated by finite element analysis (COMSOL Multiphysics with the Structural Mechanics module). **c)** Biaxial relaxation modulus $M(t) = \sigma(t)/\varepsilon$ for a constant strain in a viscoelastic material behaving as a three-parameter solid model. $M_0 = 100$ [MPa]; $M_\infty = 40$ [MPa]; $\tau_r = 10$ [s]. **d)** Sorption-induced strain $\varepsilon(t)/\lambda_i K_i C_i$ as a first-order sorption kinetics. $\tau_i = 30$ [s]. **e)** Typical responses to single analyte with $\tau_i = 5, 10, 20, 30, 60$ [s]; $M_0/M_\infty = 0.72/0.51$; $\tau_r = 22$ [s] calculated by the model in eq 7.

**Sensing System and Procedure.** The MSS chip was settled into a homemade Teflon chamber and placed in an incubator with a controlled temperature at 25.00 ± 0.02°C. The chamber was connected to gas lines consisting of four mass flow controllers (MFCs), three vials for target liquid samples, and a mixing chamber (Figure S1). Three MFCs (i.e., MFC-1, MFC-2, and MFC-3) were connected to each vial to introduce the corresponding saturated vapor. In the injection processes, the binary and ternary mixtures of homologous series with different concentrations were prepared by varying the flow rates of the MFCs (Tables S1 and S2). MFC-4 was used for diluting the vapors of binary and ternary mixtures as well as for purging by changing the flow rates every 10 s (see also Figure 2b). Total flow rate was maintained at 100 mL/min during the experiments, that is, the concentration $P_i/P_i^{o}$ of each analyte was varied in the range of 0%–30%, where $P_i$ and $P_i^{o}$ are the partial pressure and saturated vapor pressure of the *i*-th analyte, respectively. Pure nitrogen gas was used as carrier and purging gases. Data were measured with a bridge voltage of –0.5 V and recorded with a sampling rate of 10 or 20 Hz. The data collection program was designed using LabVIEW (Emerson Electric Co.).[33]

**Curve Fitting and Estimation of Parameters.** To extract the fitting parameters from pure vapors, we used least squares methods with trust region reflective algorithm using Python 3 with SciPy module according to the previous study.[33] The amplitude constant $\gamma\sigma_{sat.}$, the diffusion time constant $\tau_i$, the relaxation time constant $\tau_r$, and the ratio of unrelaxed and relaxed biaxial moduli $M_0/M_\infty$, in addition to the time $t_0$ when the first injection starts were optimized using the derived formula in eq 7 (see Results and Discussion Section). The initial fitting parameters are set as follows: $\gamma\sigma_{sat.} = 1$ [mV], $M_0/M_\infty = 3$, $\tau_i = 30$ [s], $\tau_r = 2$ [s], and $t_0 = 0$ [s]. The bounds for each parameter were set at $\gamma\sigma_{sat.} > 0$ [mV], $M_0/M_\infty \geq 1$, $\tau_i > 0$ [s], $\tau_r > 0$ [s], and $-1 < t_0 < 2$ [s].

To demonstrate the prediction of each analyte concentration, we used least squares methods with trust region reflective algorithm using Python 3 with SciPy module. Each amplitude $\gamma\sigma_i$ and $t_0$ were optimized using the derived formula in eq 7. The initial fitting parameters are $\sigma_i^{o}$, where $\sigma_i^{o}$ is $\sigma_i$ extracted from the responses to pure vapors in Table S1. The bounds for each parameter were set at $0 \leq \sigma_i \leq 1.2\sigma_i^{o}$.

## RESULTS AND DISCUSSION

**Governing Equations for Multi-component Sensing.** To derive theoretical formulations, we use a theory based on the viscoelastic behavior derived from the three-parameter solid model as follows:[32,37-39]

$$\sigma(t) + \tau_r \frac{\mathrm{d}\sigma}{\mathrm{d}t} = M_\infty \varepsilon(t) + \tau_r M_0 \frac{\mathrm{d}\varepsilon}{\mathrm{d}t}, \tag{1}$$

where $M_0$ and $M_\infty$ denote the unrelaxed (instantaneous) and the relaxed (asymptotic) biaxial moduli, respectively, and $\tau_r$ is the time constant of stress relaxation. The three-parameter solid model describes the stress ($\sigma$) / strain ($\varepsilon$) relationship in a viscoelastic solid exhibiting both viscous and elastic properties.[37,40] The behavior of the three-parameter solid model governed by eq 1 is depicted in Figure 1c. In this theory, eq 1 is directly applicable to the cantilever-type nanomechanical sensors (Figure S2; SI Text) when the coating film is significantly soft or thin film coating, i.e., $M_0 \ll M_s$ or $h_f \ll h_s$, where $M_s$ denotes the elastic biaxial modulus of the substrate; $h_f$ and $h_s$ are thicknesses of coating film and substrate, respectively.[32] Notably, a signal response of the MSS used in this study is directly proportional to the internal strain $\varepsilon_f$ in the coating film based on numerical simulations,[41] which is similar to the cantilever-type sensors.[42] Therefore, the model represented in eq 1 can also be directly applied to the MSS.[33]

In the sorption-induced nanomechanical sensing, the internal strain $\varepsilon_f$ in the receptor material at any time $t$ is modeled as a function of the concentration of analytes in the receptor material as[32,33]

$$\varepsilon_f(t) = \mathbf{\Lambda}\mathbf{C}(t), \tag{2}$$

for small volume expansion (i.e., $\varepsilon_f \ll 1$). Here, $\mathbf{C}(t)$ denotes the concentration matrix of analytes in the receptor material as a function of time and $\mathbf{\Lambda}$ is a proportional factor matrix of $\lambda_i$ corresponding to the specific volume $\nu_i$ of the $i$-th analyte (i.e., $\lambda_i = \nu_i/3$).[32,33]

In the case of typical gas sensing with nanomechanical sensors, a single rectangular injection-purge or a rectangular pulse wave-like multistep injection-purge can be considered.[33] Here, we assign the odd and even steps to the injection and purge processes, respectively (Figure 2). Let $\mathbf{C}_g$ be a matrix of the constant concentration $C_i$ of the $i$-th analyte in the gas phase during injection processes (i.e., $2n$–1 steps; $n \in \mathbb{N}$). The analyte concentrations $\mathbf{C}_g$ in the gas phase at the $n$-th step can be considered as

$$\mathbf{C}_g(t) = \mathbf{1}_A(n)\mathbf{C}_g, \quad t_{n-1} \le t < t_n, \tag{3}$$

where $\mathbf{1}_A(n)$ is the indicator function; $\mathbf{1}_A(n) = 1$ if $n$ is odd, and zero otherwise.

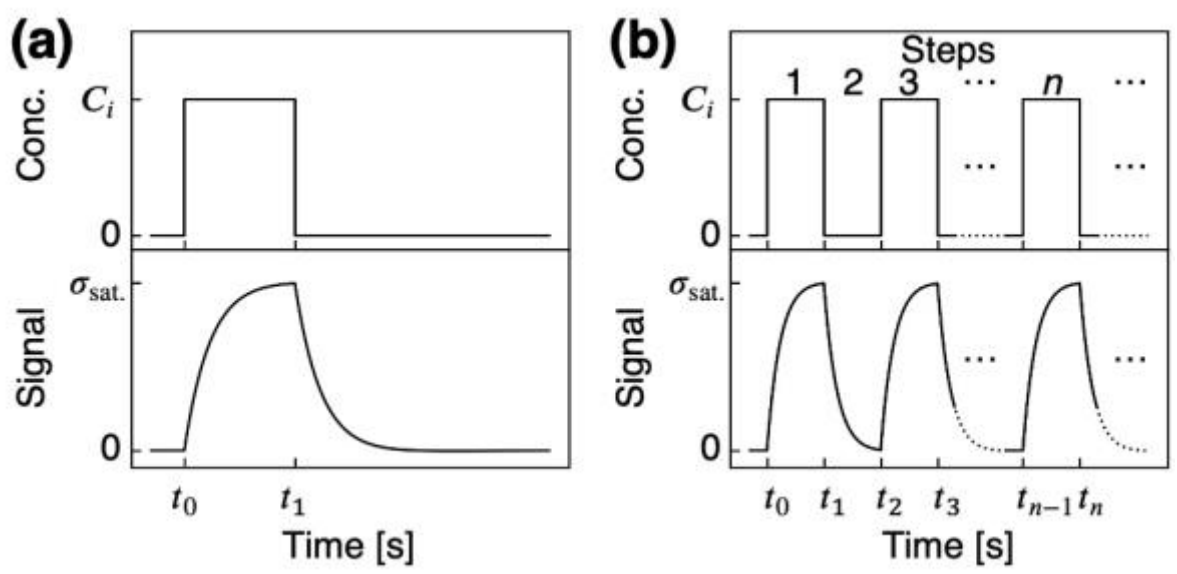

**Figure 2.** Gas injection models. **a**) A single rectangular injection-purge and corresponding typical response. **b)** A rectangular pulse wave-like multistep injection-purge and corresponding typical response.

For the derivation of the equations governing the concentration of each analyte in a receptor material coated on a nanomechanical sensor during absorption/desorption processes, we assume a first-order absorption of each analyte along with an independent reaction among analytes as illustrated in Figure 1a. Diffusion of analytes into the bulk of a coating film is generally the rate-limiting process in absorption.[31] If the diffusion of each analyte is assumed to be Fickian, the absorption rate can be approximated to be proportional to the difference between the equilibrium concentration in the receptor material and the concentration of each analyte absorbed in the receptor material $\mathbf{C}(t)$.[32,33] From eq 3, the reaction rate of the concentration of each analyte in the receptor material $\mathbf{C}(t)$ is given by

$$\frac{\mathrm{d}\mathbf{C}}{\mathrm{d}t} = \mathbf{T}_s^{-1}\left[\mathbf{K}_p\mathbf{C}_g(t) - \mathbf{C}(t)\right], \tag{4}$$

where $\mathbf{T}_s$ is a diagonal matrix of the diffusion time constant $\tau_i$ of the $i$-th analyte; $\mathbf{K}_p$ is a diagonal matrix of the partition coefficient $K_i$ of the $i$-th analyte (see Figure 1d).[32] Then, from eq 4 with eq 3, the recurrence relation can be found (see SI Text in Supporting Information). The recurrence formula can be solved, and hence the dynamic concentration $\mathbf{C}_n(t)$ at the $n$-th step can be obtained as

$$\mathbf{C}_n(t) = \mathbf{K}_p\left[\mathbf{1}_A\mathbf{I}_i - \mathbf{A}_n\right]\mathbf{C}_g, \tag{5}$$

where $\mathbf{I}_i$ is an identity matrix and a diagonal matrix $\mathbf{A}_n$ at any time $t$ of the $n$-th step is given by

$$\mathbf{A}_n(t) = \sum_{j=0}^{n-1} (-1)^j\, e^{-(t-t_j)\mathbf{T}_s^{-1}}. \tag{6}$$

Since the internal strain $\varepsilon_f$ can be assumed to be directly proportional to the concentrations of each analyte in the receptor material in eq 2 in the case of small expansion (i.e., $\varepsilon_f \ll 1$),[32,33] the dynamic stress change of nanomechanical sensors at the $n$-th step $\sigma_n(t)$ can be solved by substituting eqs 2 and 5 into eq 1 (see also SI Text) and is obtained as

$$\sigma_n(t) = M_\infty\mathbf{\Lambda}\mathbf{K}_p\left[\mathbf{1}_A\mathbf{I}_i - \mathbf{A}_n\mathbf{B} - a_n(\mathbf{I}_i - \mathbf{B})\right]\mathbf{C}_g, \tag{7}$$

with stress components in sorption kinetics $\mathbf{A}_n$ in eq 6 and in viscoelastic stress relaxation $a_n$ at any time $t$ of the $n$-th step, which is given by

$$a_n(t) = \sum_{j=0}^{n-1} (-1)^j\, e^{-\frac{t-t_j}{\tau_r}}, \tag{8}$$

where a diagonal matrix $\mathbf{B}$ is

$$\mathbf{B} = \mathbf{T}_s^{-1}\left(\frac{M_0}{M_\infty}\mathbf{I}_i - \frac{1}{\tau_r}\mathbf{T}_s\right)\left(\mathbf{T}_s^{-1} - \frac{1}{\tau_r}\mathbf{I}_i\right)^{-1}, \tag{9}$$

if $\tau_i \neq \tau_r$. In eq 7, the analytical solution clearly expresses the stress in terms of the elastic properties and stress relaxation profiles in different forms with the viscoelastic relation in eq 9. The stress $\sigma_n(t)$ given in eq 7, which is assumed to be directly proportional to the concentrations of each analyte in the gas phase, is directly proportional to the signal responses of nanomechanical sensors.[32,33] It should be noted that the stress $\sigma_{sat.}$ at the equilibrium state or a steady state of the injection process can be described as

$$\sigma_{\text{sat.}} = \lim_{t\to\infty}\sigma_n(t) = M_\infty\mathbf{\Lambda}\mathbf{K}_p\mathbf{C}_g = \sum_i \sigma_i, \tag{10}$$

where $\sigma_i$ is the stress at the equilibrium state derived from the sorption of the $i$-th analyte, which is given by $\sigma_i = M_\infty\lambda_i K_i C_i$.[33]

**Numerical Calculations of Nanomechanical Sensing.** In this section, viscoelastic material-coated nanomechanical sensing responses are numerically calculated using eq 7. One of the important features of viscoelastic behaviors in nanomechanical sensing is an overshoot as can be seen in Figure 1e.[32] If absorption of analytes occurs faster than the stress relaxation of a receptor material (Figure 1c,d), the response resulting from the absorption-induced deformation will be a signal output higher than the level of amplitude $\sigma_{sat.}$ because of the accumulation of unrelaxed stress and, subsequently, the relaxed stress (Figure 1e). In single analyte sensing,[32] the response exhibits an overshoot only if $M_0 > M_\infty$ and if

$$\tau_i < \frac{M_0}{M_\infty}\tau_r. \tag{11}$$

Figure 3a shows the numerically calculated responses to the binary mixtures with varied the stresses $\sigma_i$ derived from the $i$-th analyte. The diffusion time constants $\tau_1$ and $\tau_2$ are 5 and 60 s, respectively, with the response of an analyte 1 exhibiting an overshoot and the other not (i.e., an analyte 2), as shown in Figure 1e (see also Figure S3 in the Supporting Information for more details of other combinations of $\tau_i$). By increasing the contributions of $\sigma_1$ ($\propto C_1$), a shoulder appears in the response (e.g., $\sigma_1$ = 0.3–0.4), and then, an overshoot is observed (Figure 3a). Notably, in a binary system, even if an overshoot is observed, the response may subsequently reach a higher level of equilibrium amplitude $\sigma_{sat.}$ than that of the local maximum of the overshoot in the case of a specific balance between $\sigma_1$ and $\sigma_2$, e.g., $\sigma_1$:$\sigma_2$ = 0.7:0.3 as an example shown in Figure 3b. These trends can also be seen in the numerical calculations for mixtures of three or more analytes as shown in Figure S4 (see also supporting Python code in the Supporting Information). It is theoretically confirmed that these trends never occur in the case of a single analyte according to the previous model.[32,33]

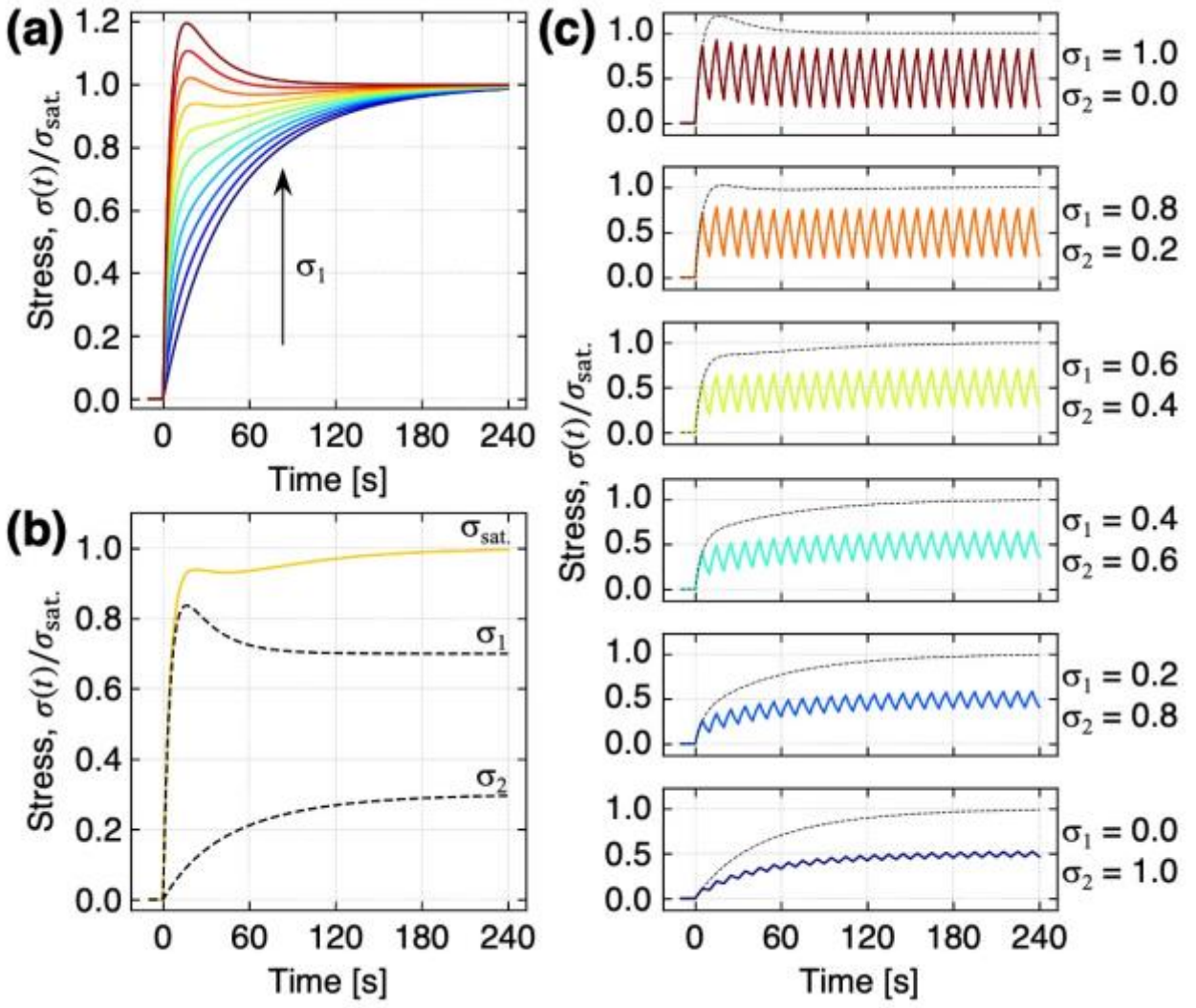

**Figure 3.** Numerical calculations of the signal responses using the derived model in eq 7. **a**) Numerically calculated responses to binary mixtures for the first injection (i.e., $n$ = 1). $\sigma_{sat.} = \sigma_1 + \sigma_2$, while $\sigma_1$ is varied from 0 to 1. $\tau_1$ = 5 [s], $\tau_2$ = 60 [s]. **b)** Response to the binary mixture with $\sigma_1/\sigma_{sat.}$ = 0.7 and $\sigma_2/\sigma_{sat.}$ = 0.3. Dashed lines correspond to the stress derived from analytes 1 and 2. **c)** Responses to the binary mixture for multistep injection-purge processes. Dashed lines correspond to the first step injection shown in (a). Other parameters are fixed: $M_0/M_\infty$ = 0.72/0.51; $\tau_r$ = 22 [s].

The present model in eq 7 is also applicable to the multistep injection-purge cycles (Figure 2b). Figure 3c shows the numerically calculated responses to various binary mixtures. If $\tau_i$ is large, analytes diffused into a receptor material during the injection process are not desorbed completely during the purge process within a limited duration, resulting in an increase in the baseline. The responses to these binary mixtures also exhibit baseline drift as shown in Figure 3c, because of the large enough values of $\tau_1$ and $\tau_2$ (see also Figure S5 for more details on other combinations of $\tau_i$; the supporting Python code for mixture of three or more analytes).

It should be noted here that a similar dynamic response of the nanomechanical sensor is often experimentally reproduced in each injection-purge cycle by repeating the cycles when the duration of each injection and purge is fixed.[11,25,41] In the case of the fixed duration $\tau$, i.e., $\tau = t_n - t_{n-1}$, eqs 6 and 8 can be simplified (see SI Text). If a number of injection-purge cycles is sufficiently large (e.g., $n \to \infty$), $\mathbf{A}_n$ and $a_n$ can be further simplified as

$$\lim_{n\to\infty} \mathbf{A}_n(t) = (-1)^{n-1}\left(\mathbf{I}_i + e^{\tau \mathbf{T}_S^{-1}}\right)^{-1} e^{-(t-t_{n-1})\mathbf{T}_S^{-1}}, \tag{12}$$

and

$$\lim_{n\to\infty} a_n(t) = \frac{(-1)^{n-1}}{1+e^{\tau/\tau_r}} e^{-\frac{t-t_{n-1}}{\tau_r}}, \tag{13}$$

respectively. From eq 7 with eqs 12 and 13, it was theoretically confirmed that each injection-purge cycle would exhibit the same dynamic response if the number of injection-purge cycles is sufficiently large, as can be seen in Figure 3c.

**Experimental Validation of the Model.** To validate the present model in eq 7, we experimentally measured signal responses of viscoelastic material-coated MSS to binary mixtures of homologous series. Since homologous series of linear alkanes and short-chain alcohols (e.g., methanol, ethanol, and propanol) tend to have similar chemical properties, each homologous series can be assumed to exhibit independent sorption behaviors.[43] It should be noted that a signal response of MSS generally correlates with the internal strain similar to the cantilever-type sensors.[33,41,42,44,45] The signal output of MSS is given by $V_{out}(t) = \gamma\sigma_n(t) + V_{out}(t_0)$, where $\gamma$ is a proportionality factor that converts stress $\sigma_n$ into an output signal of MSS and $V_{out}(t_0)$ is the signal output at $t = t_0$, i.e., baseline output (see Figure 2).[33] As viscoelastic receptor materials, polycaprolactone (PCL) and poly(vinylidene fluoride) (PVF) were coated on MSS because of their sensitivity to alkanes and alcohols, respectively. To measure the signal responses to the binary mixtures, we constructed the measurement setup as shown in Figure S1. Three MFCs (MFC-1, MFC-2, and MFC-3) were connected to each vial to introduce the corresponding saturated vapor. By varying the flow rates of MFCs in the injection processes with pure nitrogen by MFC-4, the binary mixtures of homologous series with different concentrations were prepared (Table S1).

**Table 1. Extracted Fitting Parameters of Pure Alkanes and Alcohols.**

| Sample | $\gamma\sigma_i$ [mV] [a,b] | $\tau_i$ [s] [a,b] | $\tau_r$ [s] [a,b] | $M_0/M_\infty$ [a,b] |
|---|---|---|---|---|
| *PCL/alkanes* | | | | |
| *n*-hexane | 3.05 ± 0.01 | 10.11 ± 0.20 | 1.00 ± 0.02 | 5.04 ± 0.04 |
| *n*-nonane | 2.50 ± 0.20 | 21.39 ± 3.30 | 1.63 ± 0.06 | 6.44 ± 0.58 |
| *n*-dodecane | 1.76 ± 0.13 | 22.86 ± 2.02 | 2.28 ± 0.19 | 4.76 ± 0.05 |
| *PVF/alcohols* | | | | |
| methanol | 4.65 ± 0.15 | 28.21 ± 0.93 | 1.98 ± 0.07 | 3.74 ± 0.07 |
| ethanol | 2.14 ± 0.23 | 45.81 ± 6.43 | 2.30 ± 0.04 | 3.74 ± 0.18 |
| 2-propanol | 1.22 ± 0.12 | 53.08 ± 5.83 | 1.98 ± 0.40 | 3.50 ± 0.21 |

[a] An average value ± standard deviation from six independent measurements. [b] $\gamma\sigma_i$, signal amplitude of the $i$-th analyte at the steady-state; $\tau_i$, diffusion time constant of the $i$-th analyte; $\tau_r$, relaxation time constant of coating film; $M_0/M_\infty$, the ratio of unrelaxed and relaxed biaxial moduli of coating film.

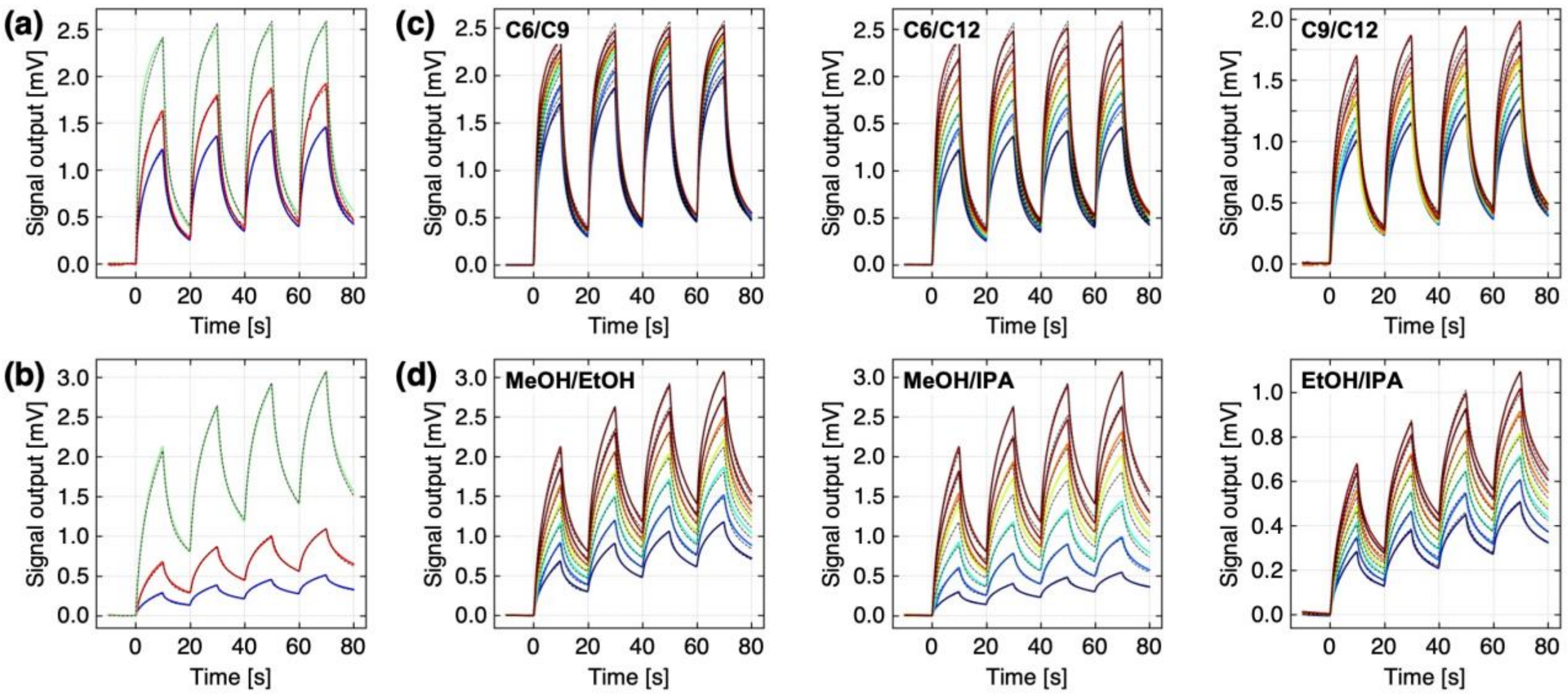

**Figure 4.** Numerically calculated dynamic responses to the binary mixtures measured experimentally by polymer-coated MSS. **a)** Responses of PCL-coated MSS to pure alkanes: *n*-hexane (**C6**; green); *n*-nonane (**C9**; red); and *n*-dodecane (**C12**; blue) with fitting results (black dashed lines). **b)** Responses of PVF-coated MSS to alcohols: methanol (**MeOH**; green); ethanol (**EtOH**; red); and 2-propanol (**IPA**; blue) with fitting results (black dashed lines). **c,d)** Responses to binary mixtures of homologous series of alkanes (c) and alcohols (d) with predicted responses based on the extracted fitting parameters (dashed lines). The color gradient from purple to red indicates the conditions from entries 1 to 7 in Table S1.

According to our previous study,[33] parameters extracted from signal responses of three to four injection-purge cycles can yield more accurate values than those extracted from a single injection. Therefore, four injection-purge response cycles of PCL-coated MSS to pure *n*-alkanes and PVF-coated MSS to pure alcohols were measured, as shown in Figure 4a,b. First, we extracted the sorption kinetic parameters and viscoelastic parameters from the responses to pure vapors, i.e., a single analyte system.[33] The fitting results are shown as dotted lines in Figure 4a,b, and extracted fitting parameters are summarized in Table 1. As expected from eqs 6–8, the sorption kinetic parameters (i.e., $\gamma\sigma_i$ and $\tau_i$) depend on the chemical properties of the *i*-th vapor, while the viscoelastic parameters (i.e., $\tau_r$ and $M_0/M_\infty$) are reasonably consistent with the vapors for each viscoelastic receptor material.

To demonstrate the predictability of the responses to multi-component vapors, the fitting parameters extracted from the responses to each pure vapor in Table 1 were used to calculate the signal responses to each binary and ternary mixture using eq 7 with theoretical concentrations controlled by MFCs (Tables S1 and S2). The comparisons between the predicted results and experimental responses of the binary and ternary mixtures are shown in Figure 4c,d and Figure 5. The predicted responses agree well with the experimentally measured responses to both the binary mixtures and the ternary mixtures, demonstrating the potential of the present model for predictive capability.

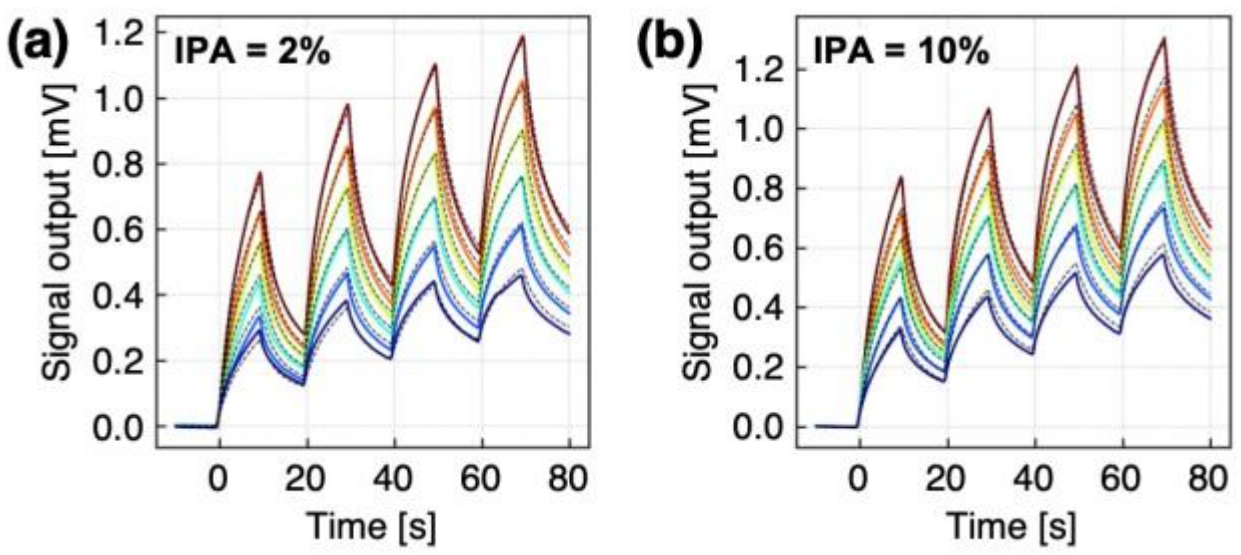

**Figure 5.** Numerically calculated dynamic responses to the ternary mixtures of alcohols experimentally measured by PVF-coated MSS. **a,b)** Responses of PVF-coated MSS to ternary mixtures of **MeOH** ($C_1$), **EtOH** ($C_2$), and **IPA** ($C_3$) with fitting results (dashed lines) based on the extracted fitting parameters ($\tau_i$, $\tau_r$, and $M_0/M_\infty$). **MeOH** and **EtOH** concentrations are varied from 0% to 10% with fixed IPA concentrations at 2% (a) and 10% (b) in Table S2. The color gradient from purple to red indicates the conditions of entries 1–6 (a) and entries 7–12 (b) in Table S2.

We also demonstrated the ability of the present model in eq 7 to predict the concentrations of each analyte (i.e., $\gamma\sigma_i \propto C_i$) in the binary mixtures using known sorption kinetic parameters (i.e., $\tau_i$) extracted from each pure vapor along with viscoelastic parameters of each material (i.e., $\tau_r$ and $M_0/M_\infty$). Figure 6 shows the fitting results and plots of predicted $\gamma\sigma_i$ from three independent measurements for each binary mixture. In the case of the binary mixtures of **MeOH** and **EtOH**, an explicit concentration dependency was observed (Figure 6d), although the predicted $\gamma\sigma_i$, especially for **EtOH**, do not linearly correlate with the theoretical vapor concentration $C_i$ because the sorption processes of each analyte in the binary mixtures are not perfectly independent. However, when the concentration of **IPA** in the binary mixture is low, the predicted $\gamma\sigma_i$ of **IPA** yields almost zero (Figure 6e,f). This tendency is particularly pronounced when the difference in $\tau_i$ is small, e.g., the binary mixtures of **EtOH** and **IPA** (Figure 6f). This may be attributed to the small contribution of **IPA** in the signal responses compared to **MeOH** or **EtOH**. Although there are some limitations, these results demonstrate the potential of the derived equation for predicting the concentrations of each analyte in the mixture.

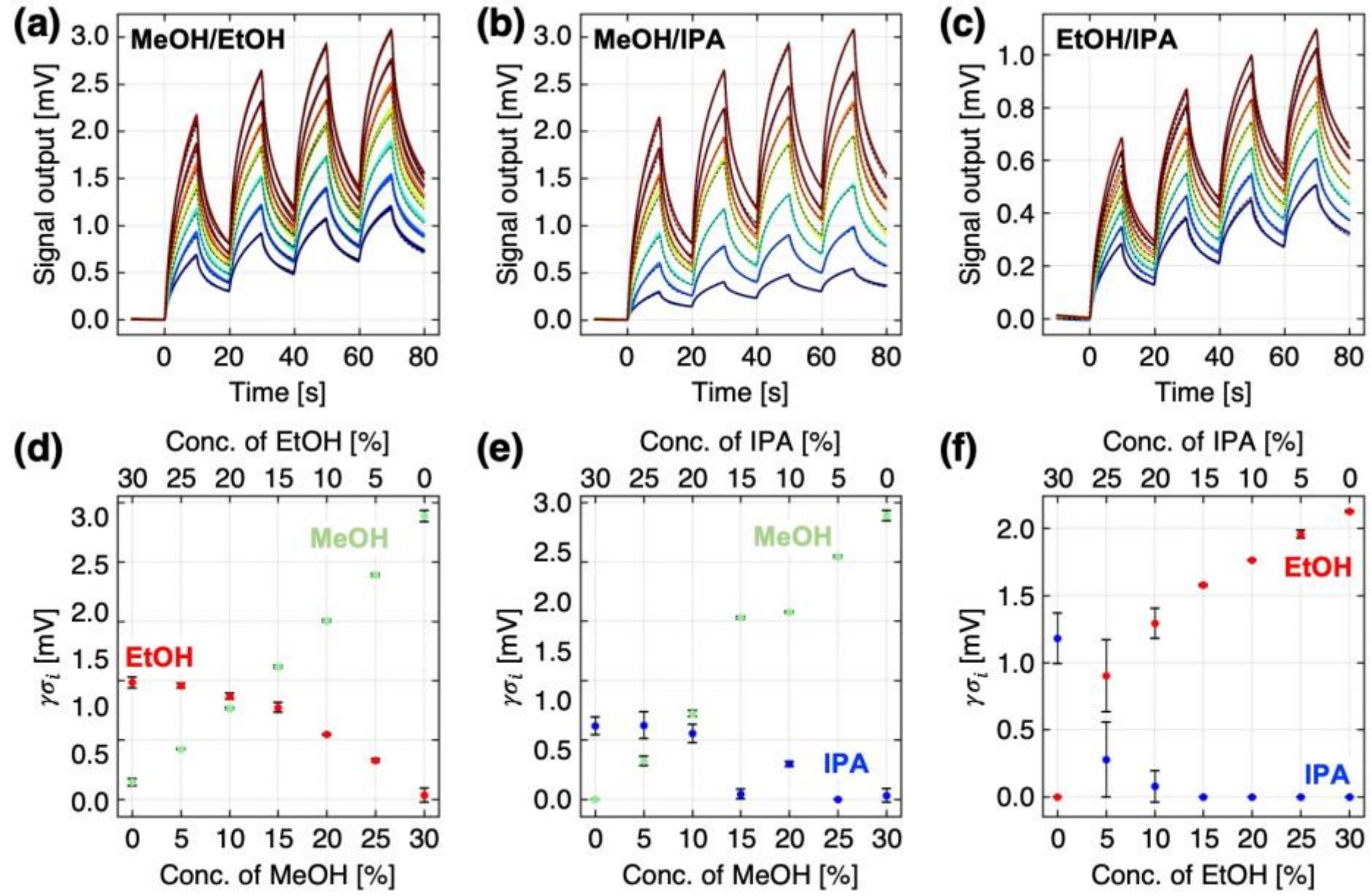

**Figure 6.** Prediction of vapor concentrations in the binary mixtures. **a–c)** Responses of PVF-coated MSS to binary mixtures of **MeOH–EtOH** (a), **MeOH–IPA** (b), and **EtOH–IPA** (c) with fitting results (dashed lines) based on the extracted fitting parameters ($\tau_i$, $\tau_r$, and $M_0/M_\infty$). Color gradient from purple to red indicates the conditions from entries 1 to 7 in Table S1. **d–f)** Plots of predicted $\gamma\sigma_i$ as a function of theoretical vapor concentrations $C_i$. Green, **MeOH**; red, **EtOH**; blue, **IPA**. Error bars are ± standard deviations.

It is noteworthy that the partition coefficient of each analyte $\log K_i$ generally yields an inverse relationship with the saturated vapor pressure $\log P_i^o$, particularly within homologous series (i.e., $\log K_i \propto 1/\log P_i^o$).[43,46] This trend implies that the partition coefficient for ethanol is higher than that for methanol. Consequently, ethanol molecules may tend to be retained more in the solid phase, while methanol molecules are likely to be in the gas phase, resulting in a convex concentration profile for ethanol and a slightly concave profile for methanol (Figure 6d). This behavior is analogous to derivations of the vapor pressure from Raoult's law, often observed in gas–liquid equilibrium.[46]

## CONCLUSION

We derived a general analytical expression that describes the dynamic responses of viscoelastic material-based static mode nanomechanical sensors to multi-component analytes. The theory includes the viscoelastic stress relaxation and sorption-induced responses with multiple analytes. Although the present model assumes that the sorption behaviors of each analyte are independent, the model is in good agreement with experimental results measured by MSS coated with viscoelastic receptor materials. The present model has the predictive capability of the dynamic responses of nanomechanical sensing, such as the trends of overshoot. Moreover, the model can be utilized for predicting each analyte concentration in the mixed vapors using the known sorption parameters extracted from pure vapors along with the viscoelastic parameters. The present model has significant potential to analyze the complex mixtures of odors as well as the analytes in the presence of interfering gases such as humidity, contributing to the development of practical artificial olfaction.

## ASSOCIATED CONTENT

### Supporting Information

The Supporting Information is available free of charge on the ACS Publications website.

Detailed derivation of eqs 7, 12, and 13, and detailed numerical calculations (PDF)
Python code for visualizing numerical calculations (Text)

## AUTHOR INFORMATION

### Corresponding Author

**Kosuke Minami** − *Research Center for Macromolecules and Biomaterials, National Institute for Materials Science (NIMS), Tsukuba, Ibaraki 305-0044, Japan; International Center for Young Scientists (ICYS), National Institute for Materials Science (NIMS), Tsukuba, Ibaraki 305-0044, Japan;*
orcid.org/0000-0003-4145-1118;
Email: MINAMI.Kosuke@nims.go.jp

### Authors

**Genki Yoshikawa** − *Research Center for Macromolecules and Biomaterials, National Institute for Materials Science (NIMS), Tsukuba, Ibaraki 305-0044, Japan; Materials Science and Engineering, Graduate School of Pure and Applied Science, University of Tsukuba, Tsukuba, Ibaraki 305-8571, Japan;*
orcid.org/0000-0002-9136-8964

### Funding Sources

This study was financially supported by a Grant-in-Aid for Scientific Research (A), JSPS, MEXT, Japan (no. 18H04168); a Grant-in-Aid for Scientific Research (C), JSPS, MEXT, Japan (no. 22K05324 and no. 25K08830); a Grant-in-Aid for Challenging Research (Pioneering), JSPS, MEXT, Japan (no. 20K20554); the Public/Private R&D Investment Strategic Expansion Program (PRISM), Cabinet Office, Japan; and ICYS, NIMS.

**Notes**

The authors declare no competing financial interest.

## ACKNOWLEDGMENT

We thank Yukiko Nakayama, NIMS, Japan for her technical assistance. K.M. acknowledges the International Center for Young Scientists (ICYS) program, NIMS, Japan.

Insert Table of Contents artwork here

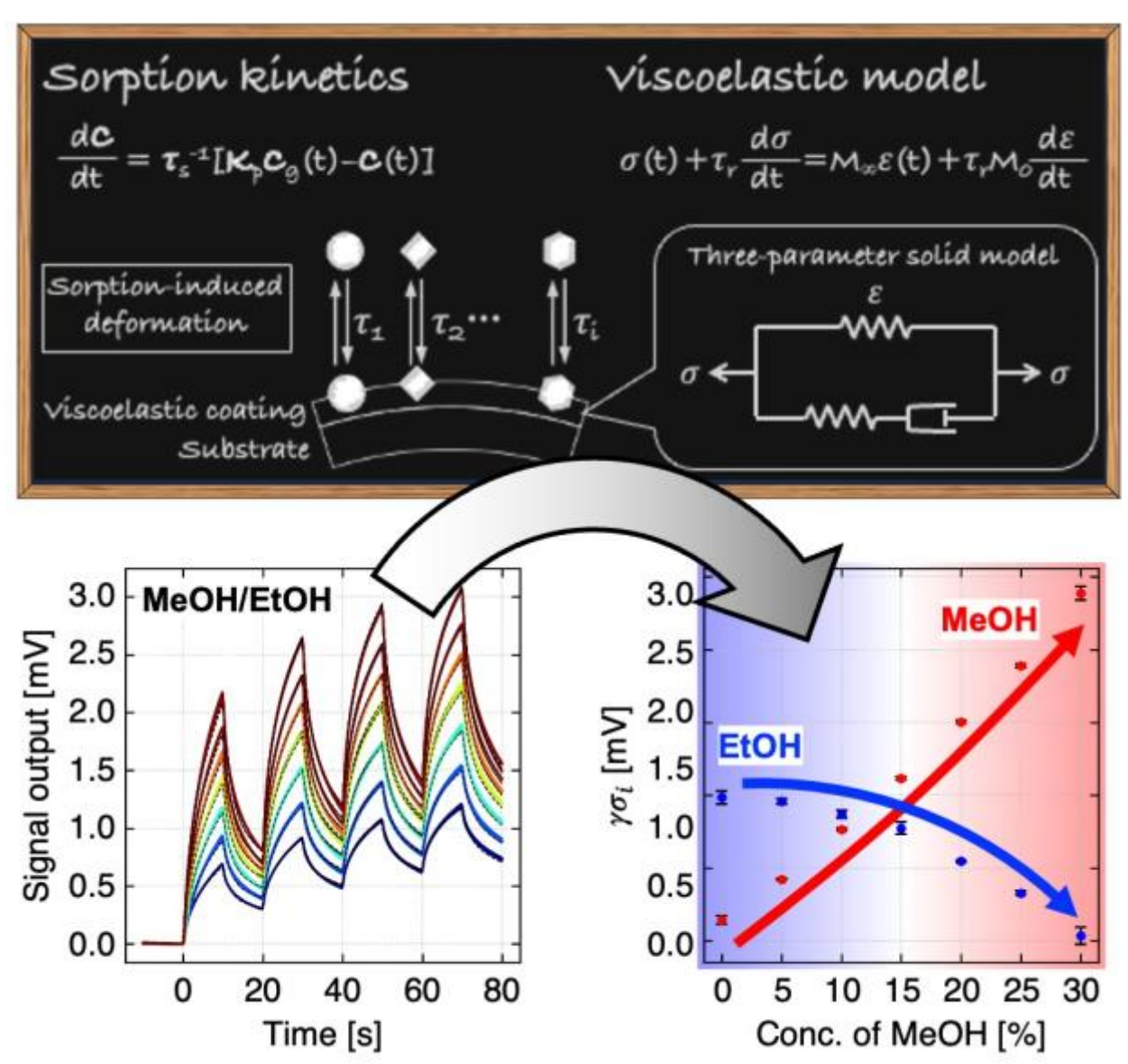